\documentclass{ieeeaccess}
\usepackage{color}

\ifCLASSINFOpdf
\else
\fi
\usepackage{supertabular}  
\usepackage{afterpage}
\usepackage{multirow}
\usepackage{longtable}
\usepackage{url}
\usepackage{color}
\usepackage{diagbox}
\usepackage{gensymb}
\usepackage{float}
\usepackage{subcaption}
\usepackage{graphicx}
\usepackage{booktabs}

\usepackage{acronym}
\newacro{AR}[AR]{Augmented Reality}
\newacro{BFT}[BFT]{Byzantine Fault Tolerance}
\newacro{BIIoT}[BIIoT]{Blockchain-based IoT}
\newacro{CPS}[CPS]{Cyber-Physical System}
\newacro{CPPS}[CPPS]{Cyber-Physical Production System}
\newacro{DBFT}[DBFT]{Delegated Byzantine Fault Tolerance}
\newacro{DoF}[DoF]{Degree of Freedom}
\newacro{DoS}[DoS]{Denial of Service}
\newacro{DPoS}[DPoS]{Delegated Proof-of-Stake}
\newacro{ECC}[ECC]{Elliptic Curve Cryptography}
\newacro{ERP}[ERP]{Enterprise Resource Planning}
\newacro{GPU}[GPU]{Graphics Processing Unit}
\newacro{GPUs}[GPUs]{Graphics Processing Units}
\newacro{HMD}[HMD]{Head-Mounted Display}
\newacro{IAR}[IAR]{Industrial Augmented Reality}  
\newacro{ICPS}[ICPS]{Industrial Cyber-Physical Systems}
\newacro{IMU}[IMU]{Inertial Measuring Unit}
\newacro{IoT}[IoT]{Internet of Things} 
\newacro{IIoT}[IIoT]{Industrial Internet of Things}
\newacro{IVR}[IVR]{Industrial Virtual Reality}
\newacro{M2M}[M2M]{Machine-to-Machine}
\newacro{MES}[MES]{Manufacturing-Execution System}
\newacro{MQTT}[MQTT]{Message Queuing Telemetry Transport}
\newacro{PBFT}[PBFT]{Practical Byzantine Fault Tolerance}
\newacro{P2P}[P2P]{Peer-to-Peer}
\newacro{PLC}[PLC]{Power Line Communication}
\newacro{PLM}[PLM]{Product Lifecycle Management}
\newacro{PoA}[PoA]{Proof-of-Activity}
\newacro{PoB}[PoB]{Proof-of-Burn}
\newacro{PoP}[PoP]{Proof-of-Personhood}
\newacro{PoS}[PoS]{Proof-of-Stake}
\newacro{PoW}[PoW]{Proof-of-Work}
\newacro{NFT}[NFT]{Natural Feature Tracking}
\newacro{RFID}[RFID]{Radio Frequency Identification}
\newacro{RSA}[RSA]{Rivest–Shamir–Adleman}
\newacro{NSA}[NSA]{National Security Agency}
\newacro{SCM}[SCM]{Supply Chain Management}
\newacro{SCP}[SCP]{Stellar Consensus Protocol}
\newacro{SDK}[SDK]{Software Development Kit}
\newacro{SDN}[SDN]{Software Defined Networking}
\newacro{TaPoS}[TaPoS]{Transactions as Proof-of-Stake}
\newacro{TLS}[TLS]{Transport Layer Security}
\newacro{VR}[VR]{Virtual Reality}
\newacro{WSN}[WSN]{Wireless Sensor Network}

\begin{document}
\history{Date of publication xxxx 00, 0000, date of current version xxxx 00, 0000.}
\doi{xx.xxxx/ACCESS.2018.DOI}
%

\title{A Review on the Application of Blockchain  for the Next Generation of Cybersecure Industry 4.0 Smart Factories}


\author{
\uppercase{Tiago M. Fern\'andez-Caram\'es}\authorrefmark{1}, \IEEEmembership{Senior Member, IEEE}\\
\uppercase{and} 
\uppercase{Paula Fraga-Lamas}\authorrefmark{1}, \IEEEmembership{Member, IEEE}}
\address[1]{Department of Computer Engineering, Faculty of Computer Science, Campus de Elvi\~na s/n, Universidade da Coru\~na, 15071, A Coru\~na, Spain. \\
(e-mail: tiago.fernandez@udc.es; paula.fraga@udc.es)}
 
\tfootnote{This work has been funded by the Xunta de Galicia (ED431C 2016-045, ED431G/01), the Agencia Estatal de Investigaci\'on of Spain (TEC2016-75067-C4-1-R) and ERDF funds of the EU (AEI/FEDER, UE). \\
Paula Fraga-Lamas would also like to thank the support of BBVA and the BritishSpanish Society Grant. 
}

\markboth
{T. M. Fern\'andez-Caram\'es, P. Fraga-Lamas: Review on Blockchain for Next Generation Cybersecure Industry 4.0 Smart Factories}
{T. M. Fern\'andez-Caram\'es, P. Fraga-Lamas: Review on Blockchain for Next Generation Cybersecure Industry 4.0 Smart Factories}

\corresp{Corresponding authors: Tiago M. Fern\'andez-Caram\'es and Paula Fraga-Lamas (e-mail: tiago.fernandez@udc.es, paula.fraga@udc.es).}


%



\begin{abstract}

Industry 4.0 is a concept devised for improving the way modern factories operate through the use of some of the latest technologies, like the ones used for creating Industrial Internet of Things (IIoT), robotics or Big Data applications. 
One of such technologies is blockchain, which is able to add trust, security and decentralization to different industrial fields.
This article focuses on analyzing the benefits and challenges that arise when using blockchain and smart contracts to develop Industry 4.0 applications. In addition, this paper presents a thorough review on the most relevant blockchain-based applications for Industry 4.0 technologies.
Thus, its aim is to provide a detailed guide for future Industry 4.0 developers that allows for determining how blockchain can enhance the next generation of cybersecure industrial applications.
\end{abstract}


\maketitle

%
\IEEEpeerreviewmaketitle

\section{Introduction}

Industry 4.0 represents the next step on the evolution of traditional factories towards actual smart factories, which are designed to be more efficient in terms of resource management and to be highly flexible to adapt to ever-changing production requirements \cite{Munera2015}.
The concepts behind Industry 4.0 are said to be first defined by the German government in 2011 \cite{HannoverFair2011,Industrie40}. Although the ideas fostered by Industry 4.0 are in part similar to the \ac{IIoT} \cite{Xu2014}, Internet Plus \cite{InternetPlus} or Made in China 2025  \cite{MadeInChina}, many industries received Industry 4.0 enthusiastically.

One of the foundations of Industry 4.0 consists in gathering as much data as possible from the diverse parts of the value chain. Such a data gathering should be performed in fast and efficient ways in order to be useful in a factory. Data collection can be carried out through systems that should allow for acquiring, storing, processing and exchanging information with devices deployed in factories or suppliers, or owned by clients.

\begin{figure*}[!bt]
\centering
\includegraphics[width=0.8\textwidth]{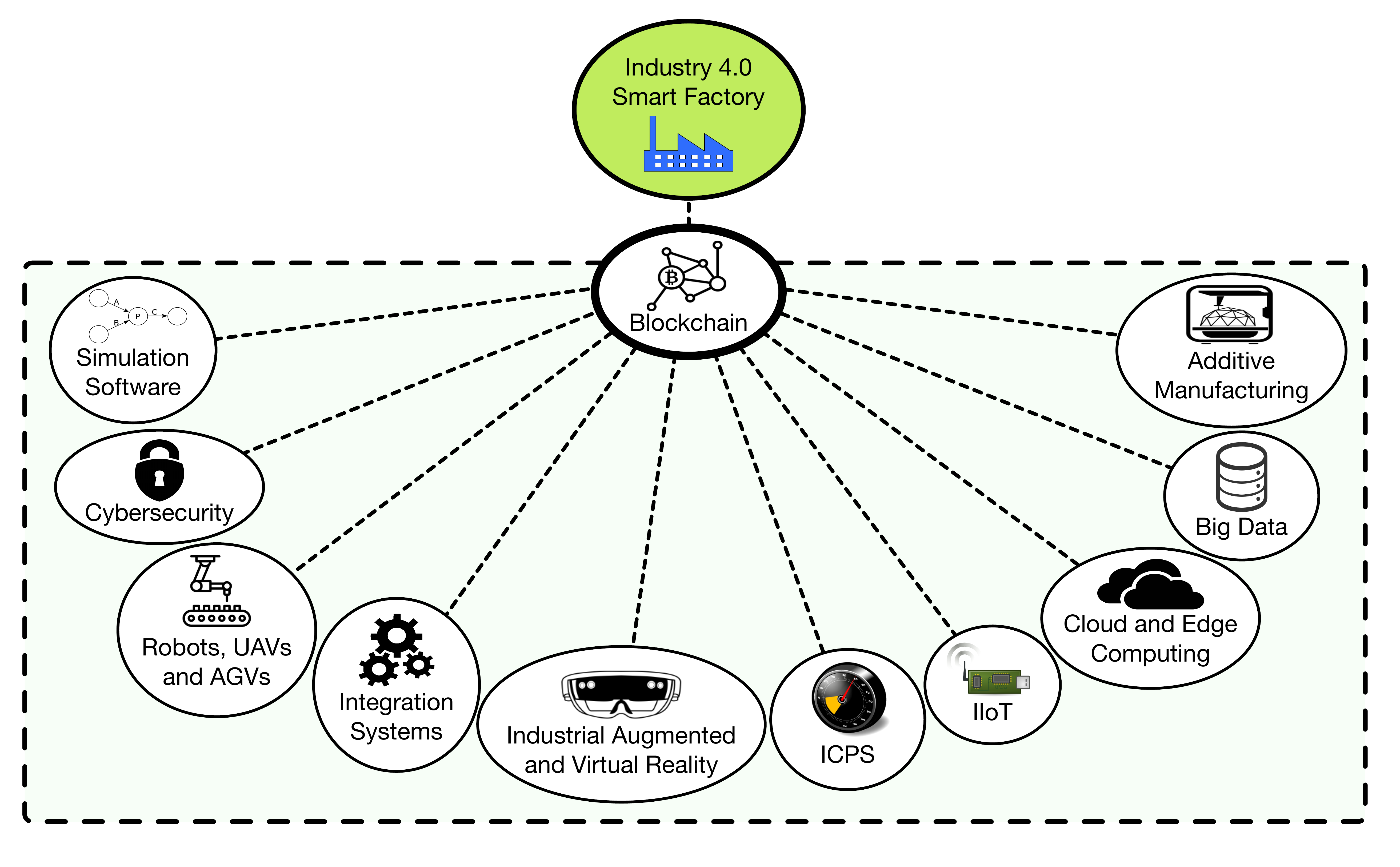}
\caption{Main Industry 4.0 technologies.}
\label{fig:tecnologiesIndustri40}
\end{figure*}

\begin{figure*}[!hbt]
\centering
\includegraphics[scale=0.2]{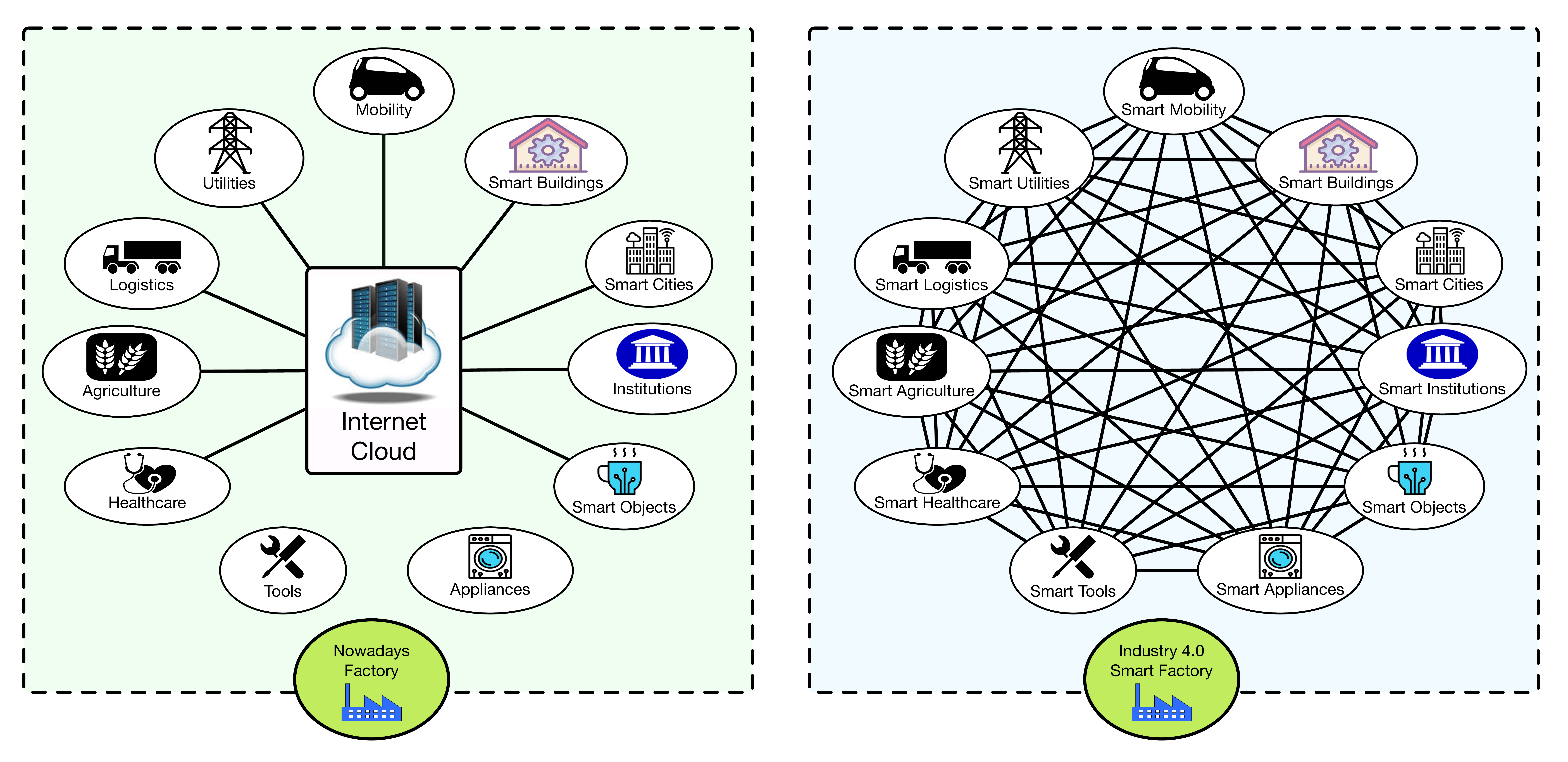} 
\caption{Evolution of the communications architecture: from modern factory to an Industry 4.0 smart factory.}
\label{fig:archi}
\end{figure*}

With the objective of providing such a level of connectivity, the Industry 4.0 paradigm proposes making use of relevant disruptive technologies that enable autonomous communications among multiple industrial devices distributed throughout a factory and on the Internet. Examples of such enabling technologies are 3D printing, \ac{AR}, \ac{ICPS}, IIoT or edge computing, which are included with other relevant Industry 4.0 technologies in Figure \ref{fig:tecnologiesIndustri40}. It is important to note that the Industry 4.0 paradigm fosters the application of such technologies to enable the evolution of the factory communications architecture from current cloud or Internet-service centered architectures, to architectures where all the entities involved in the industrial processes exchange information like in a \ac{P2P} network (Figure \ref{fig:archi} illustrates such an evolution).


One of the most promising technologies to be applied in industrial environments is blockchain, which originated from the cryptocurrency Bitcoin \cite{Nakamoto2008} and which allows for creating decentralized applications able to track and store transactions performed by a large amount of simultaneous users and devices. Blockchain can add value to different fields \cite{Swan2015}, but in this paper its ability to deal with some of the most relevant challenges currently faced by industrial scenarios is analyzed. Such challenges include:

\begin{itemize}

\item Application decentralization. It is desirable especially for Industry 4.0 applications with medium and large user and computational loads, since they rely on expensive centralized servers that are also costly to deploy and maintain \cite{Koomey2007,Middleton2015}. In addition, many industrial companies pay middlemen for outsourcing such a kind of centralized solutions.

\item It is really common to update Industry 4.0 systems (like IIoT devices) due to security issues or to upload new software/firmware. In certain scenarios the updating process requires to perform manual tasks in multiple devices spread throughout a large factory. Therefore, it is necessary to find a way to ease such tedious and inefficient tasks and distribute software updates simultaneously to as many smart devices as possible.

\item In most industries trust in the authenticity of the collected data and in the transactions performed with certain partners, service providers, manufacturers, suppliers and even governments is necessary. Due to these reasons, a technology that provides mechanisms to verify accountability and to add trust is required. Moreover, note that since trust into the collected values would be needed to replace trusted third parties \cite{Locher2018}, additional security mechanisms should be implemented, especially at a hardware level \cite{Jin2015}.

\item The data exchanged with third companies is key for some businesses, so they should be protected and anonymized. The same occurs with the data collected by IIoT devices, which should be secured and remain private to non-authorized parties. 

\item Many industrial companies also depend on closed-source code, which also increases the lack of trust, since it is actually not transparent on how the code works. Therefore, to provide trust and security, it is essential to foster open-source approaches. Nonetheless, it should be emphasized that open-source code can also suffer from bugs and exploits but, since it is verified by many developers, it is less susceptible to malicious modifications.

\end{itemize}

Most of the previous challenges can be faced by a blockchain, which acts as a ledger that stores all the performed transactions. Such transactions are validated (i.e., they are considered 'legal' according to the blockchain policy, so they are not malicious and do not create inconsistencies between users) and then added to the blockchain by computers usually called full nodes, which follow a consensus protocol (i.e., full nodes come to an agreement about whether to add the information to the blockchain).

In the case of cryptocurrencies like Bitcoin, validation tasks are performed by miners, which are computing devices that act as accountants that scrutinize every transaction. Such a validation process is essential, since it provides a decentralized solution to the Byzantine Generals' Problem \cite{Lamport82} that shows how to reach an agreement among different entities (generals) to do something (a battle plan) when such entities do not trust each other and when they only exchange messages that may be malicious (i.e., they may come from traitors).
A detailed description on the inner workings of blockchain technology is out of the scope of this paper, but the interested reader can find further information on the basics of blockchain in \cite{Puthal2018,Zheng2017,Aste2017}.

There are some recent reviews on the application of blockchain to different fields. 
For instance, in \cite{Christidis2016} an extensive description on the basics of applying blockchain to IoT is provided. Similarly, the architecture and the possible optimizations to be performed for creating and deploying blockchain-based IoT applications are detailed in \cite{TiagoBC}. Other articles focused on the application of blockchain to specific Industry 4.0 technologies like big data  \cite{Karafiloski}, to specific industries \cite{AutomotiveBC, Mondragon2018,GlucoseBC} or important characteristics like energy efficiency \cite{Mohamed2019}. 
It is also worth mentioning the systematic reviews presented in \cite{Conoscenti2016} and  \cite{Yli-Huumo2016}, which analyze the topics that papers in the literature deal with when proposing the use of blockchain. Specifically, the review presented  in \cite{Conoscenti2016} is focused on determining whether blockchain and \ac{P2P} can be employed to foster a decentralized and private-by-design IoT, while in \cite{Yli-Huumo2016} the objective is to understand the research topics, challenges and future directions regarding blockchain technology from a technical perspective.


Unlike the previously mentioned reviews, this work presents a comprehensive approach to blockchain in order to envision its potential contribution for revolutionizing Industry 4.0 technologies and confront today's challenges. This article includes the following contributions, which, as of writing, have not been found together in the literature. First, the most relevant aspects involved on the development, deployment and optimization of a blockchain are analyzed together with the advantages and practical limitations that blockchain can bring to Industry 4.0 technologies. Second, the impact of the essential parameters that influence the optimization of blockchain-based Industry 4.0 applications is studied. Third, a review on the latest and most relevant research on blockchain technologies for Industry 4.0 applications is provided. Finally, multiple recommendations are provided throughout the article with the objective of guiding future researchers and developers on some of the issues that will need to be tackled before deploying the next generation of industrial applications.


The rest of this paper is structured as follows. 
Section \ref{sec:SectionNeed} reviews different critical aspects related to the design of a blockchain-based Industry 4.0 application. 
Section \ref{sec:Applications} details the most relevant practical blockchain-based applications for diverse Industry 4.0 technologies. 
Section \ref{sec:challenges} summarizes the main challenges that arise when making use of blockchain in Industry 4.0 applications.
Finally, Section \ref{sec:Conclusions} is devoted to conclusions. 

\section{On the use of Blockchain for Industry 4.0 applications} \label{sec:SectionNeed}

\subsection{Evaluation of the need for using a blockchain in an Industry 4.0 application}


Blockchain can bring many benefits to multiple industrial fields and it can also become a very useful tool for Industry 4.0 applications. However, blockchain is not always the optimum choice to solve every problem. For instance, in private networks, traditional databases usually provide a fast and powerful tool for many applications. Therefore, before deciding whether a blockchain should be used, certain features should be recognized on an Industry 4.0 application. In fact, some authors have proposed general frameworks for determining when to make use of blockchain \cite{Lo2017}, but there are some specific Industry 4.0 application features that require a detailed analysis.

First, blockchain can be useful when decentralization is needed in an application (i.e., when such an application is enhanced by being run on a P2P network of computers instead of on a single machine). Not every Industry 4.0 technology requires decentralization, but some may benefit from it, specially when there is a centralized system that is not trusted. This happens in certain industries where, at certain point, there is no trust on specific providers, on banks and even on government agencies. Nonetheless, if there is trust among the different entities, blockchain is not needed. 

Trust is necessary when payments have to be performed. Moreover, the existence of a payment system may be used to automate many systematic tasks and thus accelerate transactions among parties. It is possible to carry out payments through traditional payment systems, but they often carry two drawbacks: they usually involve higher transaction fees than public blockchains and they have to be trusted almost blindly, without questioning their security, ethics or internal policies.

Trust and transparency are also the reasons to create public transaction logs. Such logs include certain timestamped information that may be exposed publicly (i.e., it can be scrutinized by all the entities that interact with the blockchain). 
Some Industry 4.0 applications follow this approach in a strict way and store every performed transaction to be able to carry out audits, to keep accurate traceability records or to make use of Big Data techniques or predictive analytics \cite{Cai2017,Marjani2017}.
It is worth mentioning that such features have been traditionally provided by databases, whose security is essential, especially when they are accessed through the Internet and thus exposed publicly to attacks on their availability or on the data privacy.

Another relevant feature that may be required by an Industry 4.0 application is the need for using \ac{P2P} communications to exchange data among the different parties involved in the industrial processes. This is very common in certain IIoT architectures whose nodes collaborate among them to detect specific events or to perform tasks \cite{Preden2015}.

It is important to emphasize that P2P communications are not always the best alternative to provide communications and there are alternative communication approaches that should be analyzed by an Industry 4.0 developer. For instance, in the case of resource-constraint IIoT nodes, P2P communications usually cannot be implemented efficiently due to their power consumption and the amount of demanded resources. Due to this reason, such a kind of IIoT nodes commonly route their data through gateways by using non-P2P protocols like MQTT \cite{MQTT} or by making use of a fog \cite{ZiWi} or edge computing infrastructure \cite{Cisco2012}.  

Finally, another required feature is that the distributed system should be robust. Since there are good alternatives provided by server farms or clouds, there have to be other factors to justify the use of a blockchain. The most common reason is the lack of trust in the organization that manages the infrastructure or some privacy requirements specified by a client \cite{Barhamgi}. This is essential in the case of critical infrastructures \cite{Railways2017,Thesis} and defense \cite{ReviewDefense}, whose data, due to legal and privacy concerns (e.g., certain defense data are considered strategic and therefore classified), have to be stored through trusted service providers, specially in countries where data privacy and security are not guaranteed \cite{Landau1}.

\begin{figure*}[!htb]
    \centering
        \includegraphics[width=1.98\columnwidth, height=12.8cm] {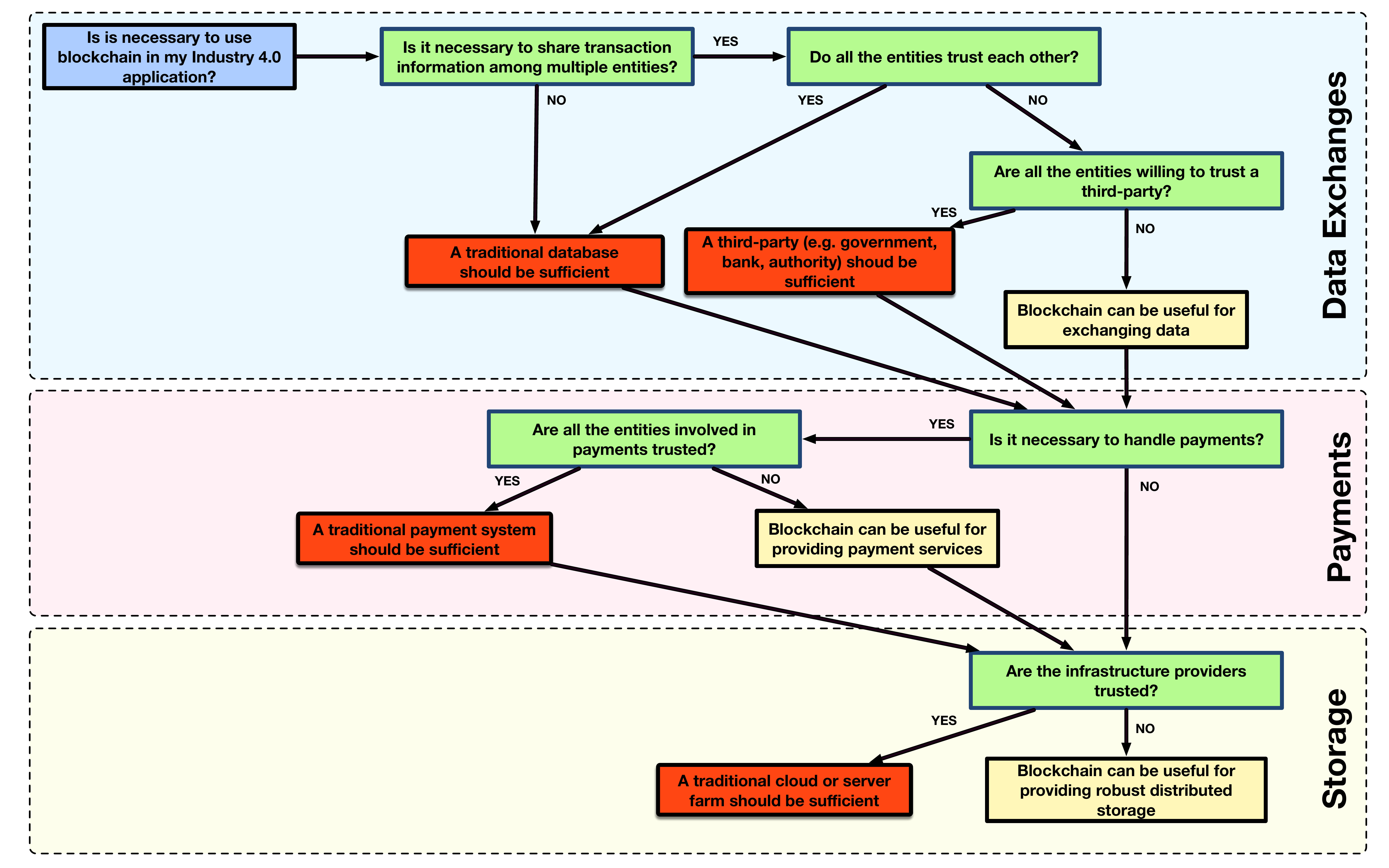}
		\caption{Flow diagram to evaluate the need of a blockchain in an Industry 4.0 application.}
        \label{figure:Blockchain_DecisionTree}
    \end{figure*}

Taking all the previous factors into consideration, Figure \ref{figure:Blockchain_DecisionTree} shows a flow diagram that can be used as a general guide for deciding when it is appropriate to make use of blockchain technologies in an Industry 4.0 application.

\subsection{Types of blockchains for industrial applications}

There exist different types of blockchains depending on how users interact with them and how data are managed/accessed \cite{TiagoBC}. Thus, the following main types of blockchain can be distinguished:

\begin{itemize}

\item Depending on access regulation, there are:

\begin{itemize}

\item Public bockchains: they do not require the approval of an entity to join the blockchain. Anyone can publish and validate transactions. Miners often receive some kind of reward for their validation work. Examples of this type of blockchain are Bitcoin or Ethereum \cite{Ethereum}. 
Public blockchains can be useful in certain industrial scenarios where a high level of transparency is necessary or where massive consumer device interaction is required.

\item Private blockchains: the participation in the blockchain is regulated by the owner. Therefore, such an owner decides on issues like the mining rewards or who can access the network.
This is the case of a blockchain like Ripple \cite{ripple}. It is important to note that, due to the existence of a single blockchain regulator, private blockchains might not be actually considered as decentralized, operating more like a close secure distributed database, which may be desirable in certain industrial environments where blockchain participants are well-known or where audits need to performed.

\item Consortium or federated blockchains: a group of owners operate the blockchain. They restrict user access to the network and the actions performed by the participants. In fact, the consensus algorithm is usually run by a pre-selected group of nodes, what increases transaction privacy and accelerates transaction validation. This can be the case of groups of industrial companies that work on the same field and that have to exchange and validate transactions: each entity may have its own validation node and when a minimum amount of nodes approves a transaction, it is added to the blockchain.
An example of consortium blockchain is Hyperledger-Fabric \cite{hyperledgerfabric}, whose ledger can be handled by multiple entities. Practical implementations of consortium blockchains can be found for financial institutions \cite{corda}, for the energy sector \cite{ewf} and for the insurance industry \cite{B3I}.

\end{itemize}

\item Depending on permissions:

\begin{itemize}
\item Permissionless blockchains: all users can perform the same actions on the blockchain, so permission management is not required. Bitcoin and Ethereum are arguably the most popular permissionless blockchains, but there are others like Litecoin \cite{litecoin}, Monero \cite{monero} or Zcash \cite{Zcash}.

\item Permissioned blockchains: they allow for controlling which users can perform transactions on the blockchain. Examples of permissioned blockchains are the ones used by Monax \cite{Monax} and Multichain \cite{Multichain}.

\end{itemize}

\item Depending on the kind of incentives:

\begin{itemize}
\item Tokenized blockchains: their transactions and incentives depend on tokens that are exchanged or given to the participants. Thus, bitcoins are used by Bitcoin and Ether is used by Ethereum.

\item Non-tokenized blockchains: they do not depend on a specific virtual currency. Examples of this kind of blockchains are Hyperledger-Fabric or RootStock \cite{rootstock}.
\end{itemize}

\item Depending on the operation mode:

\begin{itemize}
\item Logic-oriented blockchains: they enable running certain logic. Their most popular application is smart contracts (which are defined and described in the next subsection), but they can be used to execute other applications. Examples of logic-oriented blockchains are Ethereum, Hyperledger-Fabric, NXT \cite{NXT} or Counterparty \cite{Counterparty}.

\item Transaction-oriented blockchains: they are aimed exclusively at tracking digital assets (e.g., Bitcoin, Monero, Ripple).
\end{itemize}

\end{itemize}

\subsection{Smart contracts for Industry 4.0 factories}

Since blockchain can help to automate industrial processes that involve multiple companies (examples are given later in Section \ref{sec:Applications}), it is important to define the concept of smart contract: a computer program that executes agreements established between at least two parties, causing certain actions to happen when a series of specific conditions are met. Thus, when such previously programmed conditions occur, the smart contract automatically executes the corresponding clause. 
In this way, the code may translate into legal terms the control over physical or digital objects through an executable program.

Smart contract conditions are based on data that depends on external services that take data from the real world and store them into the blockchain (or vice versa). Such services are referred to as oracles. For example, an oracle could inspect records to identify whether an asset has arrived and it could write the arrival information on the blockchain. Then, the smart contract might trigger a conditional statement based on the read value and execute a block of code.

Based on the type of information collected and on the interaction with the external world, there are different types of oracles: software oracles, hardware oracles, inbound and outbound oracles, and consensus-based oracles:

\begin{itemize}
\item Software oracles handle the available online information. Examples of such an information could be the temperature of a stored product, the price of purchased parts or the traceability of the position of trucks related to logistic processes. Data come mainly from web sources and are collected by the software oracle, which extracts the needed information and pushes it into the smart contract.

\item Hardware oracles are meant to extract information directly from the physical world. For example, RFID sensors are a possible source. The biggest challenge for these hardware oracles is to report readings without sacrificing data security and to guarantee that such readings correspond to a specific physical process.

\item Inbound oracles insert information from the external world (i.e., from information sources that do not interact with the blockchain) into the blockchain (e.g., the price of an asset, which can be purchased automatically when it reaches the desired price). In contrast, outbound oracles allow smart contracts to send information to the external world (e.g., when a set of parts is confirmed to be received correctly, payment funds can be released automatically).

\item Consensus-based oracles combine different oracles to determine the outcome of an event. For instance, prediction markets like Augur \cite{Augur} and Gnosis \cite{Gnosis} make use of a rating system for oracles to confirm future outcomes and to avoid market manipulation.

\end{itemize}

\subsection{Benefits of using blockchain to enhance other Industry 4.0 technologies} 

Industry 4.0 applications share certain common issues with cryptocurrencies, since they involve many entities (e.g., IIoT nodes, operators, machines, suppliers, clients) that may not trust each other.
Nonetheless, such entities differ from cryptocurrencies in certain aspects, like the use of power constraint devices (e.g., IIoT sensors, AR glasses,  battery-operated machinery) that have to interact with the blockchain either directly (by implementing a low-consumption implementation of a blockchain client) or indirectly (through an intermediate gateway).

Despite the mentioned differences, Industry 4.0 technologies can benefit from the use of blockchain to tackle the four main challenges they face during their deployment \cite{Deloitte2015}:


\begin{itemize}

\item An Industry 4.0 factory has to deploy networks in order to connect vertically smart production systems. In a smart factory, a vertical connection is a type of connection between two entities that participate in the value chain of a product. Therefore, when such a connectivity becomes automated, information can be collected and sent automatically from the multiple systems deployed in a plant to any of the relevant parts of the value chain (e.g., to the design team or to manufacturing operators).
A blockchain can help vertical integration by providing a common trusted data or money exchange point through which the multiple smart factory entities may interact with.

\item Industry 4.0 technologies have also to be integrated horizontally, what implies that manufacturers, suppliers and clients should cooperate. Such a level of integration involves the deployment of low-latency and flexible communication networks, so a blockchain and smart contracts may become the horizontal integration mechanisms for the entities involved in Industry 4.0 processes, either for performing economic or simple data transactions.
In addition, regarding the communications between clients and companies, it can be achieved mainly through the use of IIoT devices (e.g., intelligent vehicles, smart machinery) and social networks, whose security is essential, so they may also interact through a blockchain.

\item Smart Industry 4.0 factories also require to integrate dynamically the design and engineering stages throughout the value chain. The objective of this integration is to enable fast reactions to the feedback received by the different actors that take part in the value chain. Thus, smart contracts can accelerate certain bureaucratic tasks and a blockchain can be used to carry out the mentioned interactions.

\item Integration of multiple technologies. Industry 4.0 fosters the use of different new technologies that will change the way operators interact among them and with their working environment. 
A blockchain can act as an information exchange hub whose users, which are technology independent, only have to implement the appropriate blockchain client functionality.

\end{itemize}

\section{Blockchain-based Industry 4.0 applications}  \label{sec:Applications}    

\begin{figure*}[!htb]
\centering
\includegraphics[scale=0.35]{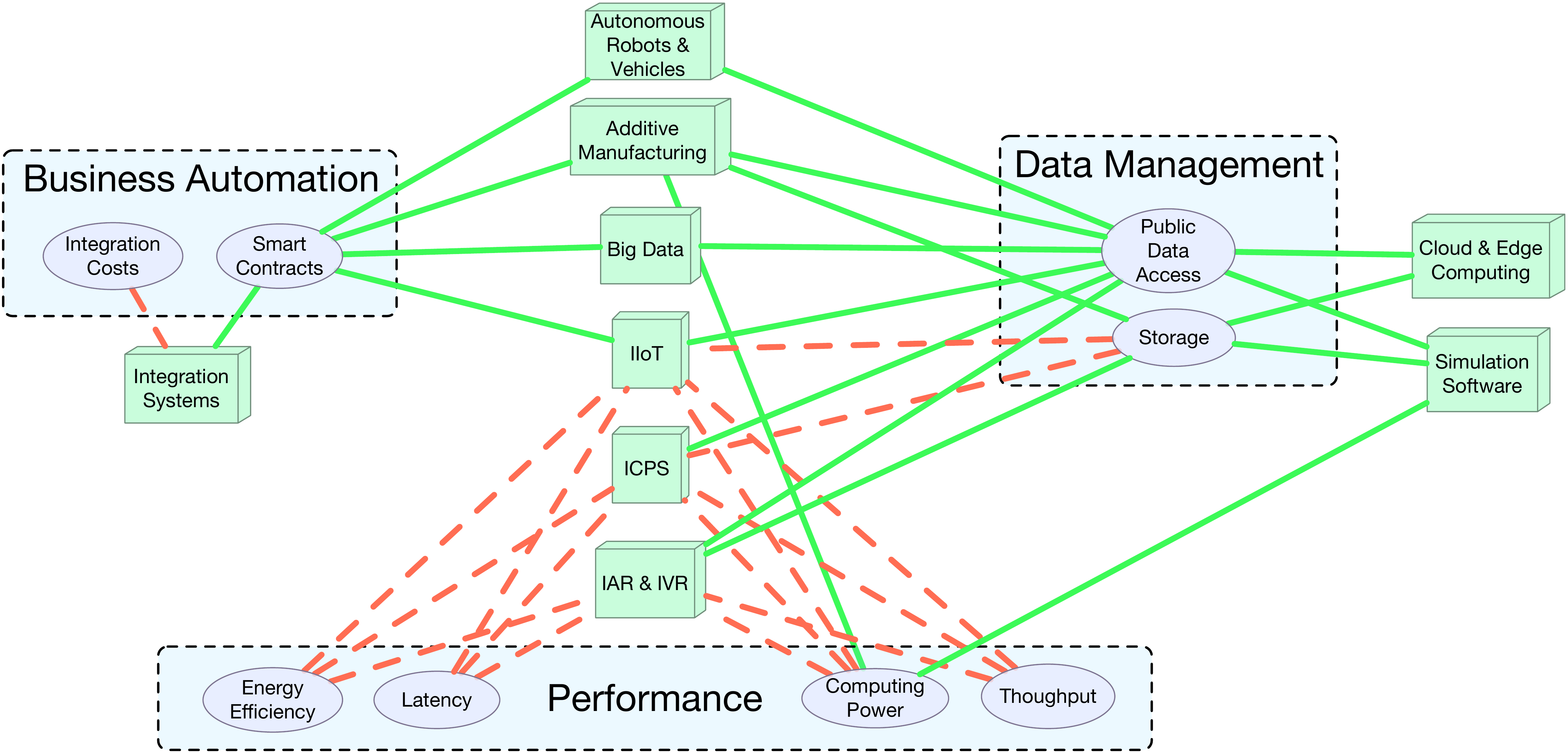}
\caption{Main benefits (green lines) and challenges (red lines) for blockchain-based Industry 4.0 applications (part 1). }
\label{fig:factors}
\end{figure*}

\begin{figure*}[!htb]
\centering
\includegraphics[scale=0.35]{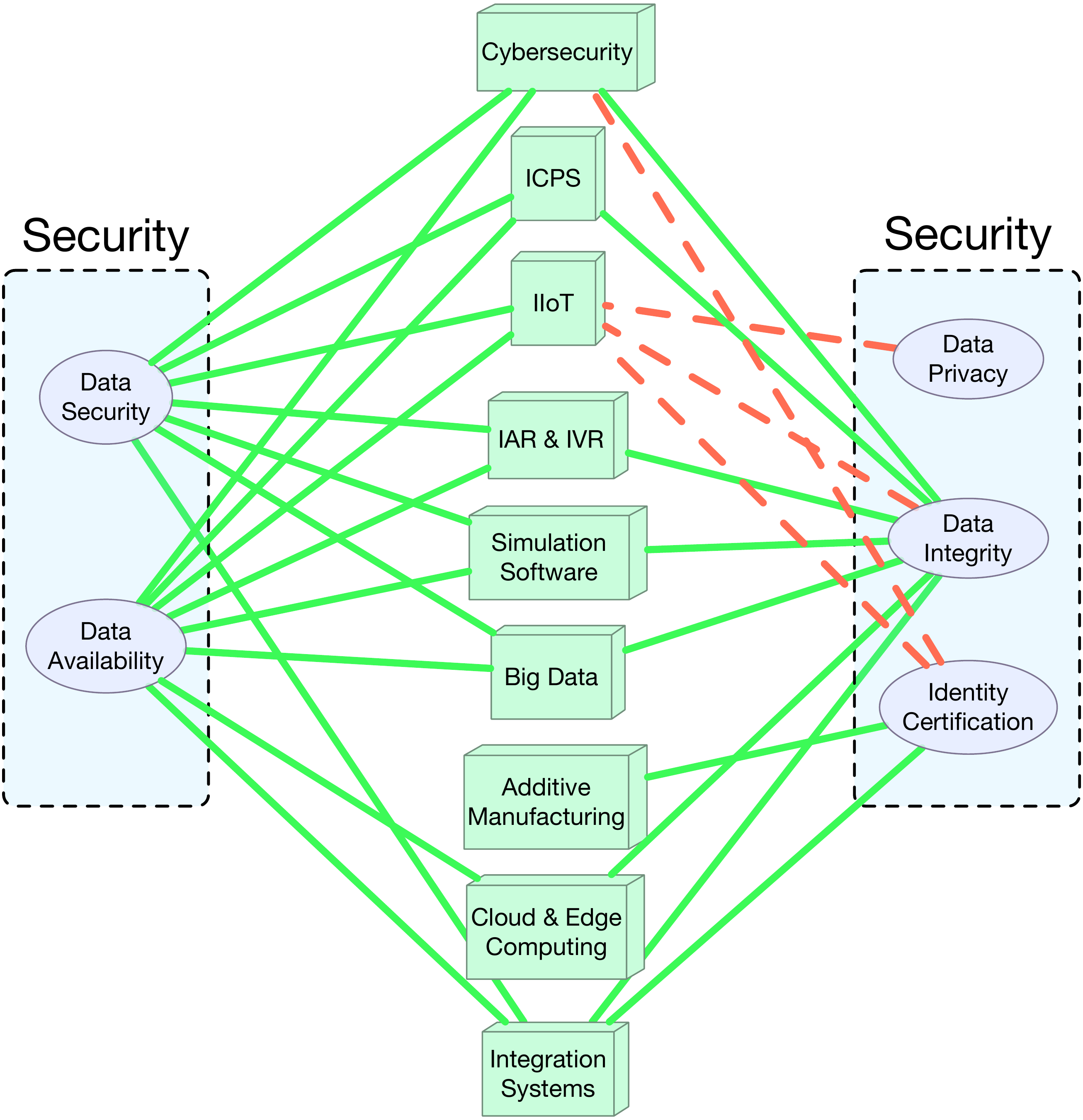}
\caption{Main benefits (green lines) and challenges (red lines) for blockchain-based Industry 4.0 applications (part 2). }
\label{fig:factors_part2}
\end{figure*}

Industry 4.0 technologies can benefit from the use of a blockchain, but its application also currently supposes a challenge in diverse aspects. Such a situation is illustrated in Figures \ref{fig:factors} and \ref{fig:factors_part2}, where it can also be observed that one feature of a technology can be improved by using blockchain, but the same feature may be still a challenge for another (e.g., the implementation of a blockchain may help cloud computing based solutions to provide redundancy for their storage needs, while, at the same time, such a local blockchain implementation is currently very difficult to replicate in IIoT nodes due their memory and computational restrictions). Therefore, when such technologies are used together in an Industry 4.0 application, it is necessary to look for a trade-off between the benefits on the use of a blockchain and its restrictions.

Figure \ref{fig:factors} also shows that certain features of a technology may be improved with the use of blockchain, but other related features may be a challenge. For instance, blockchain-based IIoT applications can increase security regarding certain aspects (i.e., data availability and communications security), but other related aspects are still a challenge (i.e., data privacy, data integrity, identity certification).
To clarify all the mentioned benefits and challenges, the next subsections analyze them and illustrate them through some of the most representative and practical  blockchain-based Industry 4.0 developments.

\subsection{IIoT}
\label{sec:IIoT}

The term IIoT refers to the use of traditional \ac{IoT} technologies \cite{Blanco-Novoa2017,intelligent_power_outlet, suarez2016,PerezExposito2017,Textile} and in industrial environments. Therefore, it implies the massive deployment of industrial sensors, actuators and machines with remote sensing/actuation capabilities \cite{Wang2015,VTEDS,Shu2017,Radon} in context aware environments \cite{Peio}. 

Blockchain can help IIoT systems to perform decentralized transactions and information exchanges during the different processing stages within a trust framework (e.g., a shared data ledger) where all the transactions are both signed and timestamped \cite{Telya2017}. An example of application is a credit-based payment scheme to support fast and frequent secured energy trading \cite{Li2018}.  
Other authors studied the advantages that an IoT-compliant blockchain can bring to the performance of the IEEE P1931.1 Real-Time Onsite Operations Facilitation (ROOF) standard \cite{Meloni2018}. In addition, blockchain-enabled IIoT developments can:

\begin{itemize}
\item Increase IIoT security. Blockchain-based applications allow for guaranteeing that the information published by a specific IIoT device is not forged or altered by malicious third parties.

\item Give public or semi-public access to IIoT data. This is important for some industries whose transparency is considered essential by certain participants (e.g., governments, supervisors, stakeholders) in order to trust a company.

\item Guarantee data access. Blockchain data are distributed among peers, so the disruption of the communications of one or more participants should not affect the access to the data. 

\item Provide a communication mechanism where multiple entities and persons interact in a standard and automated way (e.g., without any user interaction) through smart contracts (a detailed analysis on how smart contracts work is out of the scope of this paper,
but the interested reader can find a description and some examples in \cite{Christidis2016}).

\end{itemize}

Nonetheless, it must be noted that the use of a blockchain with IIoT devices is related to certain specific challenges:

\begin{itemize}
\item Data privacy. By default, all IIoT devices are identified by their public key (or its hash), so their anonymity is not guaranteed in case it is required (e.g., for instance, to avoid revealing critical data to industrial competitors). Mixing techniques increase IIoT data privacy, but in some cases they can be de-anonymized \cite{Danezis2005}.

\item Identity certification. It is important to note that when an organization sets an identity provider to authorize access to IIoT devices and other participants to the IIoT ecosystem, such a provider is also usually able to block their access. To tackle this issue some researchers propose the use of permissioned blockchains, that can manage multiple IIoT node identification \cite{Kravitz2017}, while others suggest using multichains \cite{Multichain} (where only a subset of the participants can monitor blockchain activities). 

\item High security. Some IIoT devices cannot implement complex security mechanisms due to their limited computing capabilities or their power restrictions. RSA-based cryptography may not be the most appropriate for certain resource-constraint IIoT devices, so other asymmetric cryptographic mechanisms should be studied \cite{Suarez2017}. Moreover, due to the progress of quantum computing on breaking traditional asymmetric cryptographic algorithms, post-quantum approaches should be considered \cite{Gao2018}.

\item Data integrity. It is critical for IIoT systems to guarantee that the information collected from IIoT devices has not been modified, since incorrect or malicious data can alter essential parts of an industrial system. Blockchain can provide integrity service frameworks to avoid trusting third parties \cite{Liu2017}, but such a kind of systems still has to evolve.

\item Energy efficiency. Since certain IIoT devices rely on batteries to operate, it is necessary to minimize power consumption. However, blockchains are usually power-hungry, mainly due to mining and \ac{P2P} communications (e.g., when using inefficient P2P algorithms, edge devices have to be powered on continuously and perform compute-intensive complex cryptographic operations \cite{Zhou2014,Zhang2012}). 
In addition, it must be noted that, despite possible energy consumption optimizations, current resource-constraint IIoT nodes are not able to cope with \ac{PoW}  or mining tasks in general, so they usually rely on intermediate gateways that act as proxies with the blockchain.
In the case of mining, different consensus mechanisms have been proposed to develop alternatives greener than traditional \ac{PoW} schemes (e.g., Proof-of-Stake, Proof-of-Space \cite{Dziembowski2015}), but further research is required.
For instance, some authors have already proposed alternatives like mini-blokchains \cite{Bruce2014}, which help to reduce the number of peer updates while lowering the computational requirements of a full node.
It is also worth mentioning that energy efficient hashing algorithms can help to extend the battery life of IIoT devices: although SHA-256 is probably the most popular hash algorithm due to Bitcoin, recent alternatives like Scrypt \cite{scrypt} or X11 \cite{X11} are faster, so they may reduce mining energy consumption.

\item Throughput restrictions. IIoT systems usually need to manage large amounts of transactions per second, what may be a problem for certain blockchains. For example, Bitcoin's original blockchain can reach up to 7 transactions per second \cite{Vukolic}, although it can be optimized to process more transactions (e.g., increasing block size) \cite{Courtois2014}. Therefore, throughput is an essential parameter to be considered when selecting the right blockchain for an IIoT deployment.

\item Transaction latency. It is important to note that a blockchain requires certain amount of time to process transactions. For instance, Bitcoin transaction latency follows a Poisson distribution with a 10-minute mean \cite{Nakamoto2008}, although, in practice, it is recommended to wait about an hour for the confirmation of the transaction (this occurs because five or six blocks need to be added to the chain before their confirmation). In contrast, most databases require at most a few seconds to confirm a transaction.

\item Blockchain size. Blockchains are continuously growing as IIoT devices perform transactions, what implies the use of more powerful miners. In fact, traditional resource-constraint IIoT devices are not able to manage even small blockchains. Therefore, blockchain compression techniques should be studied together with alternative approaches like mini-blockchains \cite{Bruce2014,Franca2015}.

\item Additional IIoT infrastructure. Blockchains demand infrastructure to support decentralized storage and mining. Such an infrastructure can already be implemented by using off-the-shelf hardware, but specific dedicated equipment may be needed. For example, mining hardware evolved remarkably in the last years until the development of dedicated Application-Specific Integrated Circuits (ASICs) \cite{Taylor2017}.

\end{itemize}

\subsection{Vertical and horizontal integration systems}

Horizontal and vertical integration are essential to automate data exchanges inside factories and to communicate with suppliers and clients. Traditionally, such a kind of integrations have been provided through \ac{MES}, \ac{PLM}, \ac{ERP} and IoT platforms, but Industry 4.0 demands higher integration levels, since the mentioned platforms may not be shared with other industrial partners or clients (thus requiring additional integrations, which are usually very expensive).

An example of blockchain-based horizontal integration for the power electronics industry is presented in \cite{Yan2017}. In such a paper it is proposed the use of a blockchain to develop a switching power supply in a collaborative way. Such a collaboration is performed between a manufacturer and a group of engineers. The manufacturer publishes in the blockchain (which is based on Multichain \cite{Multichain}) offers for the design of the switching supply, while the engineers read and analyze such offers, and determine whether they compete for the proposed reward. A similar system is proposed in \cite{Zhang2017}, which describes a collaborative manufacturing model for smart factories called E-Chain (Enterprise Blockchain) that allows for creating a trans-regional, trans-enterprise and trans-department industrial production service system based on blockchain.

The horizontal integration required in manufacturing and supply chain processes is a good target for blockchain-based enhancements thanks to decentralized transactions and its data management capabilities. 
In the case of a supply chain there are exchanges that require recording transactions and indicating ownership. Transaction recording also helps to keep track of the goods and also provides transparency to third parties. 
For instance, blockchain has been evaluated for its application on the supply chain of the composite material industry \cite{Coronado2018}, which manufactures products for other industries like the automotive, aerospace or railway industries.

The traceability of the supply chain is also essential in many industries, which also need it to be transparent and tamper-proof, so it can be trusted for all the involved parties. Blockchain provides such features and also adds flexibility to changing environments and regulations. 
An example of initiative for preserving traceability in industrial supply chains is detailed in \cite{Lu2017}, where it is described the development of originChain, a blockchain and smart contract based system. Such a paper also points out to the main lessons learned by the authors when developing a blockchain-based traceability application: although a blockchain provides relevant benefits, its adoption is a challenge for traditional industries due to its learning curve and the integration costs. According to the authors, the final cost of deploying a blockchain not only includes the costs related to software integration, but also the cost of the time required to understand the underlying business processes and to define precise and flexible smart contracts.

A blockchain-based consumer's supply chain can also benefit from several relevant advantages \cite{Lee2017}:

\begin{itemize}
\item It provides a tamper-proof history of product manufacturing, handling and maintenance.

\item It enables the creation of a digital identity for products, thus limiting counterfeiting and preserving ownership.

\item Thanks to smart contracts, it automates tendering among the entities and objects involved in the supply chain.

\item Since traceability can be preserved throughout the life of a product, a blockchain can be used as a tool to foster responsible consumption.
\end{itemize}

Regarding blokchain-based vertical integration, it is worth mentioning an example detailed in \cite{Backman2017}. In such a paper the authors suggest using a blockchain to negotiate automatically the reserve of resources offered by a 5G infrastructure provider. Thanks to the proposed system, mobile network and service operators can perform reservations on-demand, easily and in a dynamic way.

\subsection{ICPS}

An \ac{ICPS} or \ac{CPPS} is a system able to collect, process and store data and events to control physical processes. The different components of an ICPS are physically distributed throughout a smart factory and some of them might be on the Internet, what implies that data processing and analysis are decentralized \cite{Harrison2016,Wollschlaeger2017,Klotzer2017}, being able to make decisions in real time.

Due to the decentralized nature of an ICPS and its need for data redundancy, it seems that blockchain is a good complement for such a kind of systems. 
In fact, some researchers have already proposed the use of blockchain as the backbone of an ICPS for diverse industries. For instance, in \cite{Afanasev2018} the design of a blockchain-based CPPS for a PCB manufacturer is detailed. Other authors focused on increasing the reliability of a CPS by creating a reputation system and offering fair rewards to the diverse entities that collaborate with the system \cite{Zhao2018}.

In addition, other researchers proposed to improve manufacturing processes by decentralizing them and then creating an ICPS \cite{Isaja2018} in order to monitor and control them. In such a scenario, the researchers point out that an ICPS would benefit from the use of a blockchain to, for instance, coordinate local real-time processes, although they indicate that a problem still would have to be addressed before a commercial deployment: the serialization of transactions, which creates a performance bottleneck and that affects scalability (more computing nodes are not necessarily able to cope with higher workloads, since transactions need to be processed serially).
Moreover, it is important to note that a blockchain-based ICPS has to deal with another four limitations:

\begin{itemize}

\item Throughput and latency restrictions. Most blockchain technologies require certain time to reach consensus and confirm transactions, what may be a problem in many ICPSs that have to react in real or quasi-real time to the collected data and events.

\item Energy efficiency. Many systems that feed with data an ICPS are actually IIoT devices that may be battery operated. Therefore, it is necessary to study blockchain technologies that minimize P2P communication and consensus algorithm power consumption.

\item Computing power limitations. The IIoT devices whose data are collected by ICPS are often limited in terms of computing power, so they may not be able to make use of certain cryptographic mechanisms.

\item Storage requirements. A blockchain needs a relevant amount of storage to keep all the transactions, what may be a problem for certain resource-constraint nodes that send data to an ICPS. 

\end{itemize}

\subsection{Big Data and Data Analytics}

The ideal Industry 4.0 smart factory collects huge amounts of data from many different sources of the value chain (e.g., from plants, suppliers, logistic providers or service providers). Such an enormous amount of information is really worthy, but its processing requires using advanced Big Data techniques. In addition, Data Analytics is helpful when predicting future demands or imminent problems.

Blockchain can enhance Big Data and Data Analytics when facing three major issues \cite{Yue2017}: data collection, data trustworthiness and automated reliable data circulation. 
First, it must be noted that in the Big Data era, data comes from many scattered sources. A technology like blockchain can help to establish a common data sharing interface through which all the involved parties interact \cite{Chen2017}.

In addition, both Big Data and Data Analytics rely on reliable data to make the right decisions. Blockchain can improve data reliability by creating trust among the involved entities \cite{Abdullah2017}, securing the shared data and providing data timestamping \cite{Karafiloski}.

Regarding data circulation, it is related to the fact that information moves continuously through the Industry 4.0 ecosystem and that part of such data requires authorizations by the owners and receivers in order to be circulated. In this case, smart contracts can be really helpful, since they are able to standardize and automate data circulation, making it transparent to third parties.

\subsection{Industrial augmented and virtual reality}

An \ac{AR} device is able to display, directly or indirectly, virtual elements on top of the real-world physical environment. 
Regarding \ac{VR}, both the environment and the elements are virtual.
In the case of industrial environment, where certain industrial restrictions apply (e.g., battery life, external protection, tough lighting conditions), the terms \ac{IAR} and \ac{IVR} can be applied. 

Both AR and VR evolved remarkably in the last decade, where they proved to increase productivity \cite{Loch2016} and to be helpful in industrial design processes \cite{Cave2016,Schneider2017}, when manufacturing certain goods \cite{Boud2000, Shin2014,PracticalIAR,ReviewIAR} or for maintaining specific industrial components \cite{He2010,Alesky2014}. 

There is not much research on the application of blockchain to IAR/IVR applications, but it is clear that both technologies would benefit from the use of a blockchain regarding different aspects:

\begin{itemize}
    
\item Due to user experience (i.e., the need for fast image/video display), the most relevant AR/VR assets are stored locally, but since most AR/VR wearable devices are constraint in terms of memory and processing power, they have to rely on remote servers for data storage. Therefore, such remote servers send AR/VR data on demand upon wearable device requests. In the same way, some AR/VR applications may need to send periodic information to the central server (e.g., GPS location, sensor data) for providing certain functionality or due to traceability purposes. Thus, in these situations, a blockchain can help to share and secure the exchanged data.
				
\item A blockchain can also improve data availability. This is because AR/VR applications may need substantial bandwidth for exchanging data so, when multiple AR/VR devices communicate asynchronously with the same central server, it might become overloaded and service availability may be harmed. In this case, a blockchain can provide decentralization. For instance, the company RNDR provides distributed Graphical Processing Unit (GPU) rendering based on a blockchain \cite{RNDR}.
				
\item Enhanced data sharing and collaboration. Some organizations have proposed blockchain-based AR/VR solutions that allow for sharing digital assets in a collaborative manner \cite{Matryx} and/or in shared smart spaces \cite{Verses}.
				
\item In order for AR/VR devices to succeed commercially, they need an agile application ecosystem with an AR/VR store where users and developers can download and upload content. Certain manufacturers proposed relying on a blockchain to decentralize such an ecosystem so that to give developers freedom for creating their own storefronts and marketplaces. Examples of the mentioned ecosystems are the ones provided by Decentraland \cite{Decentraland} or VibeHub \cite{VibeHub}.

\item Certain AR/VR applications may also need to perform financial transactions with other users, which can be carried out by using cryptocurrencies or blockchain-based tokens. In this case AR/VR devices act as hardware wallets that allow for performing P2P transactions on digital tokens \cite{Lucyd}.

\end{itemize}
		
Despite the previous benefits, the use of blockchain-based IAR/IVR devices involves certain challenges: they are usually battery operated and, although they can embed really powerful \ac{GPUs}, they are restricted in terms of computing power, so they suffer from similar problems to IIoT hardware.

\subsection{Autonomous robots and vehicles}

Automation is key in Industry 4.0 applications, so the tasks susceptible to be automated may be performed by using cobots, robots or Autonomous Ground Vehicles (AGVs) \cite{Andersson2015}. Cobots are able to collaborate with humans in certain tasks \cite{Akella1999}, while robots can automate other tasks in an autonomous way \cite{Robla2017}. AGVs are used mainly for transporting or searching items throughout a factory, existing vehicles for specific industrial environments \cite{Kollmorgen,Atlas,Kiva}.
Unmanned Aerial Vehicles (UAVs) have also become very popular in the last years and they have been used in multiple industrial environments \cite{Vempati2018,Xiong2016,Harik2015,Nikolic2013}.

Blockchain can be very useful for autonomous industrial robots and vehicles, since it allows them to interact with other entities through smart contracts. Thus, blockchain-based robots and vehicles can collaborate and do business among them and with third parties in a completely autonomous way. For instance, a blockchain-based application for UAVs is described in \cite{Kapitonov2017}. In such a paper the authors detail a system based on Ethereum where UAVs interact to coordinate their air routes.
For instance, the design and some preliminary results of a UAV-based system aimed at automating the inventory of industrial items attached to active Radio-Frequency IDentification (RFID) tags is presented in \cite{droneBC}. Such a system makes use of a blockchain that receives the inventory data collected by the UAVs, validate them and make them available to the interested third parties.

There is not much research on industrial blockchain-based robots and cobots, but, in the last years, some autonomous vehicle researchers showed interest in blockchain. Most of such an interest is related to the deployment of services provided by autonomous vehicles. For instance, in \cite{Hasan2018} a blockchain-based ride-sharing service for autonomous vehicles is described. In the same way, other researchers suggested the use of blockchain for refueling, charging, parking and repairing autonomous and semi-autonomous vehicles \cite{Miller2018,Huang2018,Kang2017}.
Moreover, intelligent vehicles and robots may be rewarded according to their behavior through blockchain-based systems \cite{Yang2017,Singh2018}.

\subsection{Cloud and edge computing}

Most modern industrial companies already rely on applications deployed on local or remote cloud computing systems, which allow multiple Industry 4.0 participants to collaborate among them in an easy way. However, such a kind of system suffer from a major limitation \cite{Kshetri2017}: if the cloud is somehow affected by software problems, high workloads or attacks, the whole system may become blocked to every user. Although cloud systems enable balancing the work load and distributing part of such a load, most of them were actually not conceived from scratch as P2P systems. In contrast, blockchain-based systems are designed as distributed systems that can complement cloud computing deployments in certain aspects. For instance, in \cite{Li2017} a blockchain-based architecture for distributed cloud storage is presented. In such a system user files are divided into blocks that are encrypted, signed and uploaded to the blockchain. Then the blocks are traded among users that offer and demand cloud storage space.

Due to the previously mentioned limitations of cloud computing, alternative approaches were proposed in the last years. One of the most relevant alternatives are fog \cite{FogRA, ESP32, FogAutoID} and edge computing \cite{Dolui2017}, which is based on offloading part of the processing from the cloud to the edge of the network, what also reduces latency response.
An example of edge computing blockchain-based application is detailed in \cite{Rawat2017}, where the authors propose to use  wireless network virtualization to enhance radio spectrum utilization. In this context the blockchain is used to avoid allocating the same frequency to multiple network providers (i.e., prevent resource double-spending) and edge nodes are deployed to handle the huge amount of data expected from IoT devices, which would have to offload certain tasks due to their limited computing and storage capabilities.

\subsection{Additive manufacturing (3D printing)}

Flexibility and customization are part of the foundations of the Industry 4.0 paradigm. Both features should be provided by a smart factory without increasing the cost of the product.
In this scenario, 3D printing allows for manufacturing certain prototypes and low-volume batches at a lower price than traditional manufacturing techniques.

The following are the benefits provided by blockchain that tackle some of the main challenges of industrial additive manufacturing:
			
\begin{itemize}
    \item Decentralized supply chain. The use of a blockchain may allow for decreasing delivery time and optimizing stock management by distributing work offers from industrial consumers to producers \cite{Winkler2018}, which can also trade among them \cite{Trouton2016}. Some companies also proposed creating a global decentralized Just-in-Time Factory 4.0 by interconnecting 3D printers through a blockchain \cite{3dtoken}. Such a decentralization is also essential for sectors like Defense, where distributed additive manufacturing can enhance remarkably material readiness \cite{McCarter2017}. Moreover, such a trade can become automated through smart contracts \cite{Holland2018}, also giving transparency to third parties (e.g., auditors).
    
    \item File tracking.  Multiple designers may be involved in the creation of a 3D file, so different versions may exist, what may confuse users about which is the right version of the file to work with. In this scenario, a blockchain can act as a distributed database for the file hashes, which can be sorted by their timestamp in a single repository so that file use can become synchronized. For example, the additive manufacturing solution Digital Factory from LINK3D \cite{link3d} is already using blockchain for file integration.
    
    \item Distributed computing. Most 3D printing design process need computer power to perform tasks like when creating the nesting before launching the additive manufacturing processes or when preparing a design's slicing. In these cases, powerful computers can be used or cloud computing platforms, but also a blockchain that can distribute the computing calculations among multiple peers (maybe involving economic incentives) in order to accelerate the 3D printing process.
    
    \item Increased trust. Additive manufacturing involves interacting with multiple actors and suppliers, which may not be necessarily trustworthy. A blockchain can help by improving transparency and validating the origin of the manufacturing files \cite{GE2018}.
    
    \item Intellectual property protection. Plagiarism is a concern in the field of additive manufacturing, specially for industrial companies, since once a design is delivered to one user, it can be easily replicated infinitely. To avoid such a problem, in \cite{Holland2018} is proposed a whole additive manufacturing supply chain that includes the use of smart contracts to create licensing agreements so that the licensee can only print a specific amount of copies of the licensed component.
\end{itemize}

\subsection{Cybersecurity}

Intra and inter-connections are key in Industry 4.0 applications, so it is necessary to protect the systems involved in such connections. In addition, security is essential for industrial critical systems, which have been widely targeted by cyber-attacks in the last years \cite{Hur2017}, affecting both complex industrial systems and simpler physical access systems \cite{conferenceRFID,Reverse2017,Methodology2017}.

As it has been already mentioned in previous subsections, since most blockchain technologies use by default secure public-key cryptosystems and hash algorithms, data security is enhanced respect to other solutions that provide it optionally. In addition, private and consortium blockchains can restrict user access, reducing the number of possible attackers. Furthermore, since the data on the blockchain are distributed, although one participant may be under attack, the information can be available through other nodes, thus guaranteeing data availability. An example of blockchain-based information sharing framework designed to face cybersecurity challenges is described in \cite{Rawat2018}.

Nonetheless, it is important to note that blockchains, like other types of distributed systems, are prone to Sybil attacks \cite{Douceur2002}, which can alter the proper behavior of the system. Moreover, a blockchain can be boycotted by a group of miners when they become a majority (which is not difficult in small blockchains), being able to block certain transactions due to economic or ideological reasons. Therefore, although blockchain technology can increase Industry 4.0 application security, its use entails certain security issues that should be identified and addressed \cite{Conti2018}.

\subsection{Simulation software}

The information collected by an Industry 4.0 factory can be used to model the behavior of all the entities involved in the production system (e.g., machines, operators, products) by making use of simulation software. Such a software can determine the current state in the real-world of the factory (this is related to the concept of Digital Twin \cite{Qi2018}) and then predict future events, suggest mitigation measures to avoid problems, or suggest improvements to reduce costs or to improve quality.
Specifically, blockchain technology can help simulation software by:

\begin{itemize}
    \item Collecting data from multiple sources in a distributed repository whose availability is therefore improved thanks to the use of the redundant information provided by different nodes.
    
    \item Verifying data authenticity, thus removing part of their uncertainty. In this way, predictions on certain parameters may become more accurate and give managers a better picture on the state of a factory at a specific time instant.
    
    \item Distributing computational tasks. Blockchain is able to distribute tasks and calculations among different nodes in order to accelerate simulations. In fact, some cryptocurrencies like Gridcoin \cite{gridcoin} have already proposed rewarding with tokens peer collaboration in solving mathematical problems in a distributed way.
    
    \item Providing enhanced Simulation-as-a-Service and decentralized co-simulation functionality \cite{InfiniteFoundry}.
			
\end{itemize}

\section{Main challenges of the implementation of blockchain into Industry 4.0}
\label{sec:challenges}

Despite the benefits provided by blockchain technologies, their development and deployment in Industry 4.0 applications suppose significant challenges  that require further research:

\begin{itemize}
 
 \item Scalability. The architectures selected to support blockchain-based Industry 4.0 applications would have to withstand the significant amount of traffic that such applications usually generate. Such an amount may be a problem for  traditional centralized cloud-based architectures, which in the last years have evolved towards architectures that tend to provide the most basic services close to the point where they are physically required, as it happens with fog and mist computing architectures \cite{Preden2015}.

 \item Cryptosystems for resource-constraint devices. Many devices that operate in Industry 4.0 factories (e.g., sensors, actuators, tools) have very limited computational resources (i.e., memory and processing power), so they struggle with modern secure public-key cryptography schemes \cite{Li_authentication_2017}. Although most blockchains make use of public-key cryptosystems based on Elliptic Curve Cryptography (ECC), which is generally lighter that traditional Rivest-Shamir Addleman (RSA) when compared at the same security level \cite{Suarez2018}, such a kind of cryptography is still power hungry. In addition, the industries that need to keep data secure for the medium and long-term should be aware of the threat of post-quantum computing \cite{NSA} and look for energy-efficient quantum-safe algorithms.

 \item Consensus algorithm selection. Since the consensus algorithm is essential for the proper working of the blockchain, it has to be selected carefully. It is important to note that it is not necessary to use the most egalitarian (and idealistic) consensus mechanism, which would consist in giving to all the miners the same weight when voting, since in certain blockchains (e.g., public blockchains), such a mechanism would be prone to Sybil attacks and then a single entity would be able to control the whole blockchain. Besides traditional \ac{PoW} consensus algorithms like the one used by Bitcoin, there are others like Proof-of-Stake, Proof-of-Space, Proof-of-Activity, \ac{PBFT} \cite{Castro1999}, Sieve \cite{Cachin2016}, Proof-of-Burn or Proof-of-Personhood \cite{Borge2017}. In addition, it is also worth noting that mining, although it is very useful in public blockchains, it is not indispensable for every scenario, so the computational effort and energy consumed by the consensus algorithm can be drastically reduced.

 \item Privacy and security. As it was detailed in Section \ref{sec:IIoT}, data privacy, identity certification and data integrity are still issues that need to be addressed properly, specially for resource-constraint devices.

 \item Energy efficiency, throughput and latency. These issues were previously described in relation to IIoT and CPSs, but they can be extrapolated to other blockchain applications. In the case of energy efficiency, the use of mining, inefficient P2P protocols and computationally complex cryptographic algorithms have an impact on the energy consumption in every scenario, although such factors are critical when battery-operated devices are used. Regarding blockchain throughput and latency, they are influenced by way the consensus algorithm works and how blocks are added to the blockchain. In fact, both factors usually increase latency and reduce remarkably the throughout respect to, for instance, traditional database systems, so it may be difficult to provide real-time responses to events.
 
  In the case of Ethereum, currently every single node must process every single transaction that goes through the network \cite{Mohanty}.
  Nevertheless, Ethereum is making a steady progress towards scalability with the development of on-chain solutions like sharding \cite{Sharding} or off-chain ones like Raiden \cite{Raiden} and Plasma \cite{Plasma}.   
Sharding allows nodes and transactions to be divided into smaller partitions called shards with their own state and transaction history.
Therefore, certain nodes would process transactions only for certain shards allowing a higher throughput.

Unlike sharding, Raiden proposes to scale the Ethereum network using off-chain transactions.  It is the Ethereum's version of Bitcoin's Lightning Network \cite{LightningNetwork}. It allows a collection of nodes to establish bidirectional payment channels to facilitate near-instant, low-fee and scalable microtransactions, without directly transacting with the Ethereum blockchain.
Recently, a simplified version of this solution called $\mu$Raiden \cite{microRaiden} enables to make micropayments through unidirectional payment channels.

Another off-chain solution is Plasma, a proposed framework for enforced execution of smart contracts. Plasma creates blockchains that follow a tree hierarchy by allowing for the creation of child  blockchains enforced by their parent. There are three main implementations Minimal Viable Plasma (MVP), Plasma Cash and Plasma Debit.


 \item Required infrastructure. The use of blockchain technologies requires to deploy specific hardware infrastructure, like additional storage or mining hardware. Moreover, the foreseen high amount of data traffic generated by P2P communications require communications infrastructure and interfaces able to support the estimated load.

 \item Management of multi-chains. The proliferation of blockchains may require some companies to support several of them simultaneously. For instance, a company may handle its financial transactions using Bitcoin, while smart contracts are executed on applications that rely on Ethereum. Therefore, solutions would have to be designed and implemented to use different blockchains at the same time.

\item Interoperability and standardization. Currently most companies develop their own blockchain solutions, but interoperability among them is necessary in many scenarios in order to achieve a seamless integration. Some entities like the IEEE are working on specific standards aimed at guaranteeing interoperability in diverse fields. For instance, as of writing, two relevant initiatives are being driven by the IEEE Standards Association \cite{IEEEStandards} in relation to blockchain: one related to build consensus on optimizing clinical trials and enhancing patient safety, and another one for driving collaboration on advancing blockchain adoption within the pharmaceutical industry.

\item Regulatory and legal aspects. Besides the previously mentioned technological challenges, it is also essential to pay attention to new regulations and laws that are being developed by governments and regulatory agencies. For example, the European Union launched in February 2018 its Blockchain observatory and forum, whose main objective is to map key initiatives, monitor developments and develop common actions at an EU level \cite{EUobservatory}.

\end{itemize}

\section{Conclusions} \label{sec:Conclusions}

Industry 4.0 is a paradigm that is changing the way factories operate through the use of some of the latest technologies.
One of such technologies is blockchain, which has been successfully used for cryptocurrencies and which can enhance Industry 4.0 technologies by adding security, trust, immutability, disintermediation, decentralization and a higher degree of automation through smart contracts.

This article provided a detailed review on the benefits that blockchain can bring to the main Industry 4.0 technologies, as well as their current challenges. After providing a general methodology to determine whether the use of a blockchain is an appropriate choice for implementing an Industry 4.0 application, the most relevant industrial blockchain-based applications for every Industry 4.0 technology were studied, as well as their main challenges. In this way, this article provides a guide for the future Industry 4.0 application developers in order to determine how blockchain can enhance the next generation of cybersecure industrial applications.

 

\begin{IEEEbiography}[{\includegraphics[width=1in,height=1.25in,clip,keepaspectratio]{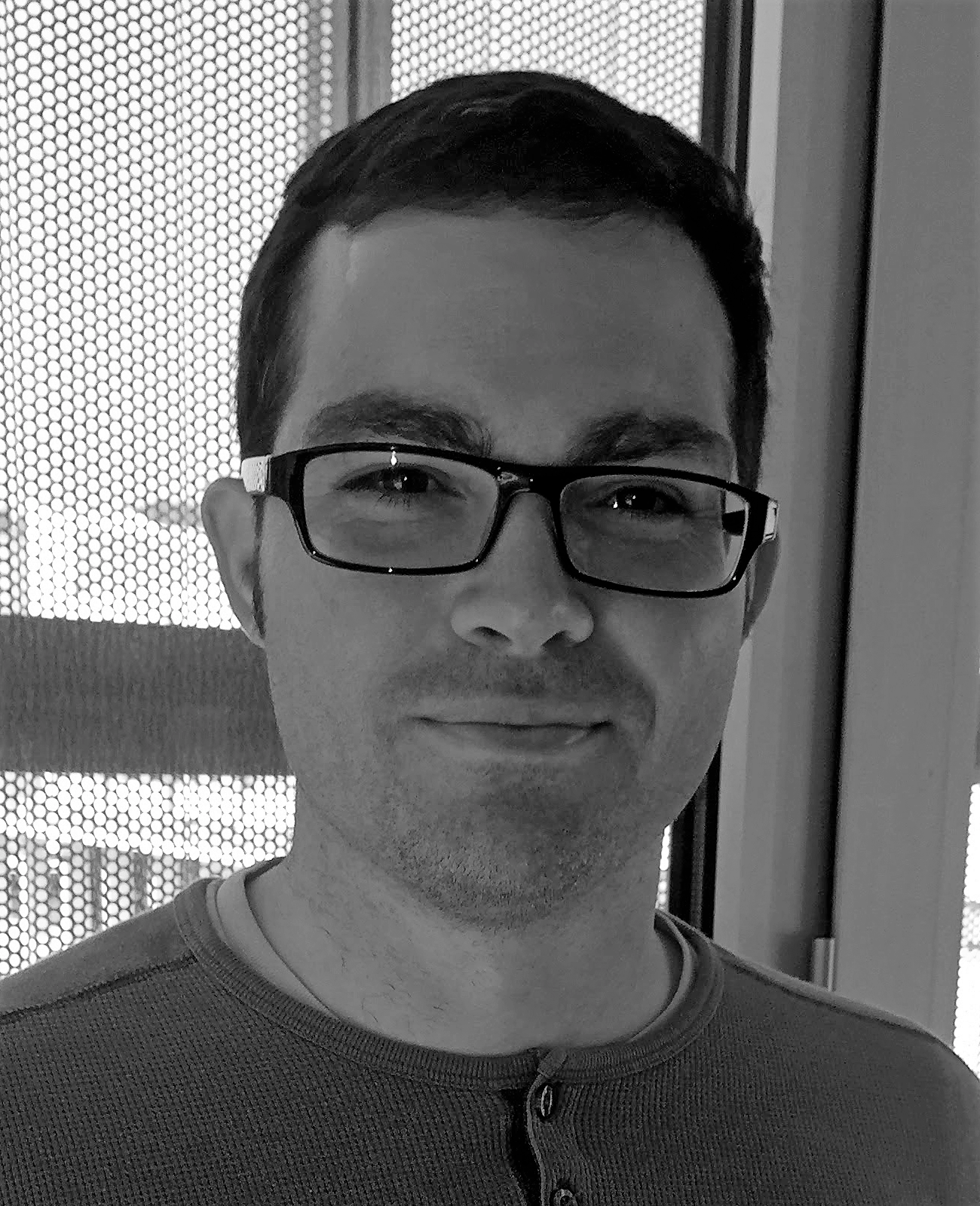}}]{Tiago M. Fern\'andez-Caram\'es} (S'08-M'12-SM'15) received his MSc degree and PhD degrees in Computer Science in 2005 and 2011 from University of A Coru\~na, Spain. Since 2005 he has worked as a researcher and professor for the Department of Computer Engineering of the University of A Coru\~na inside the Group of Electronic Technology and Communications (GTEC). His current research interests include IIoT/IoT systems, RFID, wireless sensor networks, Industry 4.0, blockchain and augmented reality. 
\end{IEEEbiography}

\begin{IEEEbiography}[{\includegraphics[width=1in,height=1.25in,clip,keepaspectratio]{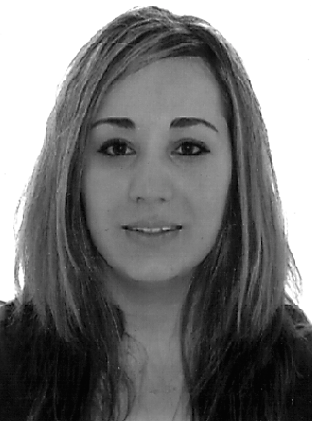}}]{Paula Fraga-Lamas}(M'17) received the M.Sc. degree in Computer Science in 2008 from University of A Coru\~na (UDC) and the M.Sc. and Ph.D. degrees in the joint program Mobile Network Information and Communication Technologies from five Spanish universities: University of the Basque Country, University of Cantabria, University of Zaragoza, University of Oviedo and University of A Coru\~na,  in  2011 and 2017, respectively. Since 2009, she has been working with the Group of Electronic Technology and Communications (GTEC) in the Department of Computer Engineering (UDC). She holds an MBA and postgraduate studies in business innovation management (JMC of European Industrial Economy), sustainability (CSR) and social innovation (INDITEX-UDC Chair). She is co-author of more than fifty peer-reviewed indexed journals, international conferences and book chapters. Her current research interests include wireless communications in mission-critical scenarios, Industry 4.0, Internet of Things (IoT), Augmented Reality (AR), blockchain, RFID and Cyber-Physical systems (CPS).  She has also been participating in more than twenty research projects funded by the regional and national government as well as R\&D contracts with private companies. 
\end{IEEEbiography}

\EOD
\end{document}